# Estimating seed bank accumulation and dynamics in three obligate-seeder Proteaceae species

Meaghan E. Jenkins,[1] David A. Morrison,[1,2*] and Tony D. Auld [3]

[1] Department of Environmental Sciences, University of Technology Sydney, Westbourne Street, Gore Hill NSW 2065, Australia
[2] Section for Parasitology (SWEPAR), Department of Biomedical Sciences and Veterinary Public Health, Swedish University of Agricultural Sciences, Uppsala 751 89, Sweden; David.Morrison@bvf.slu.se
[3] Plant Ecology Unit, Climate Change Science Section, Department of Environment & Climate Change (NSW), PO Box 1967, Hurstville NSW 2220, Australia; Tony.Auld@environment.nsw.gov.au

**Abstract**

The seed bank dynamics of the three co-occurring obligate-seeder (i.e. fire-sensitive) Proteaceae species, *Banksia ericifolia*, *Banksia marginata* and *Petrophile pulchella*, were examined at sites of varying time since the most recent fire (i.e. plant age) in the Sydney region. Significant variation among species was found in the number of cones produced, the position of the cones within the canopy, the percentage of barren cones produced (*Banksia* species only), the number of follicles/bracts produced per cone, and the number of seeds lost/released due to spontaneous fruit rupture. Thus, three different regeneration strategies were observed, highlighting the variation in reproductive strategies of co-occurring Proteaceae species. Ultimately, *B. marginata* potentially accumulated a seed bank of ~3000 seeds per plant after 20 years, with ~1500 seeds per plant for *P. pulchella* and ~500 for *B. ericifolia*. Based on these data, *B. marginata* and *B. ericifolia* require a minimum fire-free period of 8–10 years, with 7–8 years for *P. pulchella*, to allow for an adequate seed bank to accumulate and thus ensure local persistence of these species in fire-prone habitats.

## 1 Introduction

The ecological effects of fire are many and complex, and it is unlikely that they will ever be well understood for all organisms and ecosystems (Burrows *et al.* 1999). However, if there are characteristics of species that indicate limits to possible fire intervals for their existence, then these limits provide a starting point for determining the frequency distribution of fire intervals in which the species can occur (Gill & McCarthy 1998). Analysis of the responses of individual species or guilds to fire in terms of population processes can also provide the basis for the development of predictive models (Cowling *et al.* 1987). Functional classification of species offers a powerful means of ordering knowledge into generalisations, as well as a guide for determining research priorities and applying fire management to species for which data are scarce (Keith 1996, Pausas *et al.* 2004). In particular, species that are abundant and common can be used as indicators, based on which educated decisions could be made with respect to fire management practices to ensure informed conservation of biodiversity in these ecosystems.

---
* Corresponding author.

Serotiny, the canopy storage of seed in closed woody fruits for a prolonged period, is a trait evident in fire-prone vegetation throughout the world (Bond 1985, Cowling & Lamont 1985, Lamont *et al.* 1991), and it is a common characteristic found in a number of important families occurring in fire-prone habitats of Australia (Enright *et al.* 1998). The ecological significance of serotiny seems clear: where conditions for seedling recruitment are best immediately after a fire occurs (i.e. competition for light, moisture and nutrients are probably at a minimum), cued release of canopy-stored seeds after fire maximises the age, and therefore accumulation of seed store, of the population by the time of the next fire (Enright *et al.* 1996). Some seeds may germinate between fires, but the general absence of seedlings and very young plants in unburnt vegetation has lead to the suggestion that the majority of seedlings recruited between fires do not survive (Zammit & Westoby 1988). Furthermore, Bradstock & O'Connell (1988) observed the existence of new cohorts of *Petrophile pulchella* and *Banksia ericifolia* seedlings in long unburnt populations (greater than 20 years since the last fire) but short-term mortality of these was high. Therefore, seedling recruitment in these species seems to be most successful after fire when conditions are more favourable for growth and survival, illustrating the advantage of serotiny in fire-prone habitats.

Proteaceae species provide an excellent opportunity for exploring the evolutionary significance of fire-adaptive reproductive traits and mechanisms that promote co-existence of species (Cowling *et al.* 1990). For example, 76% of *Banksia* species store their seed bank in their canopy and release their seeds after fire (i.e. serotiny) (Cowling & Lamont 1985). Therefore, they are ideal for the study of seed bank dynamics due to their canopy storage of seeds, which is a distinct advantage for study compared to those species which store their seed bank in the soil, as accumulation and annual contribution can be easily assessed (Bradstock & Myerscough 1981; Witkowski *et al.* 1991). However, canopy seedbank species often represent a relatively small proportion of the vascular flora in a plant community. So, they may not be good indicators of other species with transient or persistent soil seed banks, except perhaps for the effects of fire frequency, where they may be regarded as more sensitive since they have no residual seed bank unless fires are patchy. Fire frequency is the most critical aspect of fire regimes for such plants, because fire repeatedly interrupts processes such as fruit production and growth that maintain the capacity of the population to persist and regenerate (Keith 1994).

In this study we aimed to quantify the pattern of accumulation of seeds in the canopy seed bank of three common, co-occurring, obligate-seeding (i.e. fire-sensitive) Proteaceae species of eastern Australia: *Banksia ericifolia*, *Banksia marginata* and *Petrophile pulchella*, using a chronosequence analysis. We then estimated the minimum fire-free interval allowable to ensure their persistence in fire-prone communities.

## 2. Materials and Methods

*2.1. Study Species and Areas*

All three Proteaceae species are shrubs to 5 m that are abundant in the understorey in sclerophyll vegetation in the Sydney region. The seed bank dynamics of these three obligate-seeder species were measured at 13–20 sites per species during March–August 2001 (Table 1), covering a wide range of times since the last fire (TSLF), which should equate with plant age for these fire-sensitive species. The sample sites were located on the southern edge of Sydney, in Royal and Heathcote National Parks, Garawarra and Dharawal State Recreational Areas, and within the Cordeaux Catchment area. Furthermore, four additional sites were sampled in Ku-ring-gai Chase National Park to the north of Sydney, approximately 60 km from the previous sites, for *B. ericifolia* and *P. pulchella*, thus allowing for comparisons to be made regarding spatial variation and differences within and between populations of the same species.

**Table 1.** Number of plants sampled for each species at each site. An asterisk indicates a site where individual cones were also sampled.

| Site | Time since last fire (years) | *Banksia ericifolia* | *Banksia marginata* | *Petrophile pulchella* |
|---|---|---|---|---|
| *Southern sites* | | | | |
| Heathcote 1 | 3 | 30 | 30 | 30 |
| Dharawal 1 | 3 | 30 | | |
| Dharawal 2 | 3 | | | 30 |
| Cordeaux 1 | 5 | 30 | | 30 |
| Royal 1* | 7 | 30 | 30 | 30 |
| Royal 2 | 7 | 30 | 30 | 30 |
| Cordeaux 2 | 8 | 30 | 30 | |
| Cordeaux 3* | 8 | 25 | 30 | 30 |
| Cordeaux 4* | 10 | 30 | | 30 |
| Dharawal 3 | 10 | 30 | 30 | 30 |
| Dharawal 4* | 10 | | 30 | |
| Cordeaux 5* | 15 | 30 | 29 | 30 |
| Dharawal 5 | 16 | 30 | | 30 |
| Heathcote 2* | 19 | 30 | 29 | 30 |
| Garawarra 1* | 21 | 30 | 30 | 30 |
| Garawarra 2 | 21 | 30 | 30 | |
| Heathcote 3* | 24 | 30 | 30 | 30 |
| Heathcote 4 | 24 | 30 | 30 | 30 |
| *Northern sites* | | | | |
| Salvation Loop 1 | 7 | 25 | | 20 |
| Salvation Loop 2 | 8 | 25 | | 20 |
| Ku-ring-gai Chase | 9 | 25 | | 25 |
| Mt Ku-ring-gai | 10 | 20 | | 20 |

Sites were selected to cover a representative range of ages available with the following selection criteria: (1) the time since the last fire was accurately known from fire-history maps obtained from the New South Wales National Parks and Wildlife Service; (2) the site contained at least one of the three study species but preferably more; (3) the population appeared to be even aged (i.e. all individuals had been killed by the most recent fire); (4) sites with more than 50% exposed rock were not used, due to the patchy nature of fires in these habitats; and (5) care was taken to sample a reasonable distance (usually approx. 5–10 m) from roads, walking and maintenance trails, and any other disturbances, to avoid edge effects, i.e. possibly higher growth rates of plants on roadsides and potential fire patchiness in these areas (Lamont *et al.* 1994).

*2.2. Sample Technique*

A target of 30 individuals of each species was sampled at each site (Table 1), using randomly selected line transects to chose the individuals; however, where species were in low abundance all available individuals were sampled. The Cordeaux 3 site (Table 1) showed an apparently varied-age population of *B. ericifolia*, suggesting that there had been a patchy fire; so a sample size of 25 individuals for this species was used and unusually large individuals (i.e. those presumably not burnt and killed in the last fire) were avoided. Replicate sites were sampled for each time since the last fire (TSLF) (Table 1) if several fires occurred in the same year, or where the area burnt by the last fire was sufficiently large to allow geographically separate populations to be sampled. For the samples taken in the northern sites the sample size was reduced to 20–25 individuals (Table 1), as this sample size had already given reliable results concerning the data for the southern sites.

All three Proteaceae species have their fruits aggregated into confructescences, which we have called "cones". For the two *Banksia* species each fruit is a woody follicle on the cone, while for the *Petrophile* species it is a nut enclosed by a woody bract on the cone. The numbers of new (i.e. produced in 2001) and mature (i.e. pre-2001) cones were counted on each sampled individual. Furthermore, the percentage of barren cones, i.e. cones that had produced no follicles, was calculated for both *B. marginata* and *B. ericifolia*. Canopy volume data were obtained from Jenkins *et al*. (2005), measured for the same populations at the same time.

All three study species have a distinctive growth habit of annually produced whorls of shoots, and this can be used to accurately age the plants (Jenkins *et al*. 2005). It can also be used to approximately estimate the time of production of the cones. Therefore, an estimate of each cone's age was taken by counting from the newest cone (i.e. the current season's) downward towards the base of the plant, ageing each cone on each whorl in succession. Some cones needed to be grouped into age classes due to difficulties encountered with accurately ageing older cones.

For selected southern sites (Table 1), a random sub-sample of the individuals had 3–4 cones of each age class (i.e. whorl position) examined for the total number of follicles or bracts, number of open follicles/bracts and number of closed follicles/bracts. The cones were harvested for *P. pulchella* and *B. marginata*, due to difficulties in obtaining measurements for these species in the field, while the data obtained for *B. ericifolia* were all recorded *in situ*. The number of follicles was not calculated on the current season's cones for *B. marginata*, as many cones had not fully developed at the time of sampling.

The total accumulated seed bank for each population was then estimated per plant as follows:
Seed bank = average number of cones per plant × proportion of cones fertile × average number of fruits per fertile cone × proportion of non-open fruits × number of seeds per fruit.

The number of seeds per fruit was obtained from previous studies: *B. ericifolia* – 1 seed per follicle (George 1981), *B. marginata* – 1.78 seeds per follicle (Vaughton & Ramsey 1998) and *P. pulchella* – 1 seed per bract (D.A. Morrison pers. obs.).

*2.3. Data Analyses*

For the relationship between TSLF age (as the independent variable, x) and the total number of cones per plant (as the dependent variable, y), mean values per population were calculated and least-squares regression analyses were then conducted on the logarithm-transformed data (Minitab 2000). After estimating the x-value for a y-intercept of zero in the regression, this was set to x = 3 for both *Banksia* species but left at x = 0 for *P. pulchella*. For the relationship between the number of new (i.e. 2001) cones produced per plant and plant canopy volume, mean values per population were calculated and least-squares correlations were then performed (Minitab 2000). For the relationship between TSLF age (as the independent variable) and the total accumulated seed bank (as the dependent variable), the single estimates for each population were analysed by least-squares non-linear curve-fitting of a sigmoid function (Raner 2001). This function was chosen as a simple heuristic tool, rather than with any specific biological model in mind.

The effects of plant age on the counts of the number of fertile/barren cones in the current (i.e. 2001) and older seasons at the different sites were analysed by contingency chi-square analyses (Minitab 2000). The effects of plant age on the number of follicles/bracts per cone in the current and older seasons were analysed by 1- or 2-factor analyses of variance using a general linear model (Minitab 2000). The effects of cone age (i.e. whorl position) on the number of open and closed follicles/bracts per cone were analysed by 1-factor analyses of variance (Minitab 2000).

## 3. Results

All three species displayed an increase in the total number of cones with TSLF age (Fig. 1), as would be expected for any serotinous species. *B. marginata* displayed the highest level of cone accumulation, with *B. ericifolia* and *P. pulchella* producing approximately half as many cones as *B. marginata*. No cones were produced in the 3-year-old populations for either *Banksia* species, and very few for *P. pulchella*. The fitted regression equations predict that the average time for production of the second cone on a plant would be 8.3 years TSLF for *B. ericifolia*, 4.5 years for *B. marginata* and 4.3 years for *P. pulchella*. These can be considered as minimum estimates of the earliest useful contributions to an accumulating seedbank. However, it is likely to be an under-estimate for *B. marginata* because no 5-year-old site was sampled, and this affects the shape of the relevant part of the fitted regression. Indeed, a linear regression fits the *B. marginata* data just as well as does the non-linear one (this is not true for either the *B. ericifolia* or *P. pulchella* data), and this predicts that second cone production on a plant would occur at an average of 6.7 years TSLF, which we consider to be a more realistic estimate for this species.

For both *B. ericifolia* and *P. pulchella* the older sites had much greater variation in the number of accumulated cones, both within and between populations, than did *B. marginata* (Fig. 1). Furthermore, the number of new cones produced was positively correlated with the current volume of the plant canopy for all three species: *B. ericifolia*, $r = 0.89$, $P < 0.001$; *B. marginata*, $r = 0.74$, $P < 0.001$; *P. pulchella*, $r = 0.98$, $P < 0.001$. The slightly poorer correlation value and the lesser variability for *B. marginata* are both consistent with the more-linear increase in cone production noted above. There were no detectable differences in fruit production between the northern and southern sites of the same TSLF age for either *B. ericifolia* or *P. pulchella* (Fig. 1).

The growth-position of the newly produced cones varied between the species (Fig. 2), although all three species preferentially produced the cones on younger whorls. *P. pulchella* almost always produced cones on 2-year-old shoots, while *B. ericifolia* generally preferred 6–7-year-old shoots, but with considerable latitude in the position, and *B. marginata* preferred 3–4-year-old shoots. These position preferences may be related to the age required for first flowering in the species, and this position can also be used to determine a minimum age for the older cones.

Both *Banksia* species produced substantial numbers (from 10–50%) of cones with no apparent reproductive value (Fig. 3). For the accumulated cones (i.e. those produced in previous years), neither species showed any significant trend with plant age in the proportion of fertile/barren cones (*B. ericifolia*, $\chi^2 = 8.91$, $P = 0.113$; *B. marginata*, $\chi^2 = 6.08$, $P = 0.299$). However, the current (i.e. 2001) season's crop for *B. ericifolia* had fewer fertile cones than did the older seasons ($\chi^2 = 44.39$, $P < 0.001$). This may indicate that older infertile cones drop off the plants, and their number is thus under-estimated when counted several years later, or it may indicate a change in plant behaviour with age or particular seasonal conditions. Furthermore, the proportion of fertile/barren cones varied significantly between the sites for the current season of *B. ericifolia* ($\chi^2 = 26.52$, $P < 0.001$), with many more fertile cones produced at the youngest site (7 years TSLF). This presumably reflects a behavioural difference between the first major fruiting season of a population and subsequent seasons.

The average number of fruits per cone was calculated for the *Banksia* species using only the cones that produced follicles, due to the high percentages of barren cones, and using all cones for *P. pulchella* (Table 2). Both *Banksia* species showed a significant difference in the average number of fruits produced per cone with respect to plant age (Table 3), with a greater number of fruits per cone at 15–20 years TSLF (Table 2). *P. pulchella* did not show this pattern. On the other hand, *P. pulchella* generally produced significantly fewer fruits per cone in the current (2001) season than in previous seasons (Tables 2, 3), while *B. ericifolia* showed no difference.

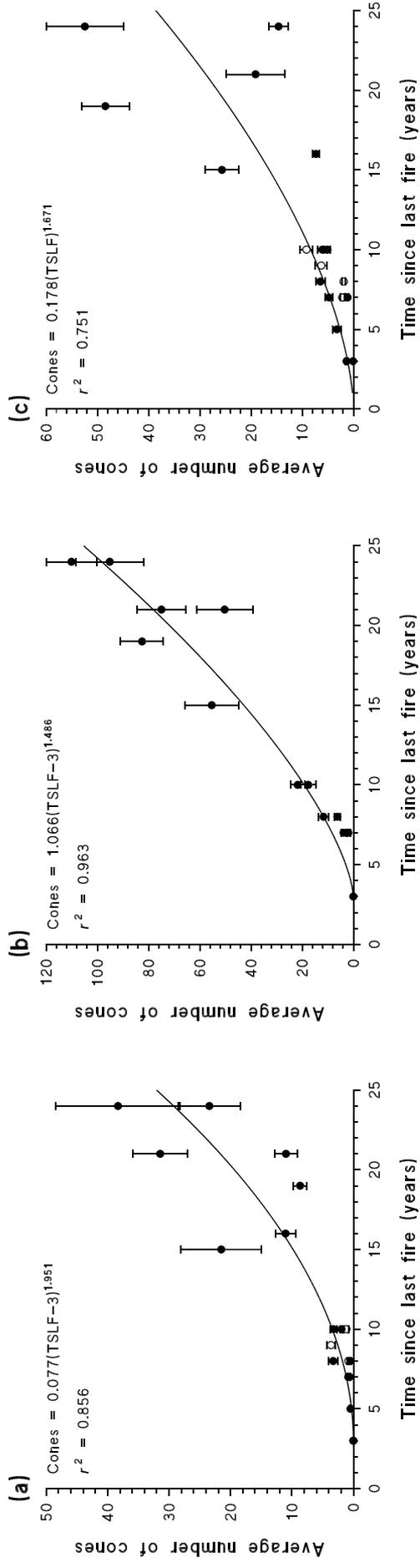

**Figure 1.** Relationship between the average number of cones per plant and time since the last fire (TSLF) for (a) *B. ericifolia*, (b) *B. marginata* and (c) *P. pulchella*. Each symbol represents a population average with standard errors, showing the southern (filled symbols) and northern (open symbols) sites, along with the predicted values (solid line) of the best-fit regression.

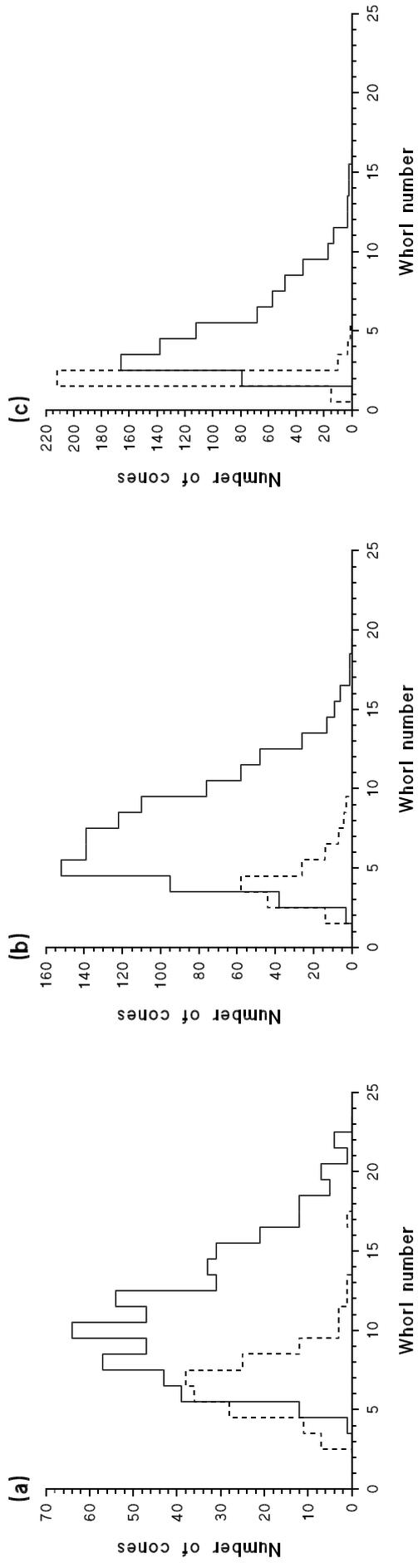

**Figure 2.** Frequency histograms of the growth-whorl position (≈ shoot age) of the current season's cones (dashed lines) and the older cones (solid lines) for (a) *B. ericifolia*, (b) *B. marginata* and (c) *P. pulchella*.

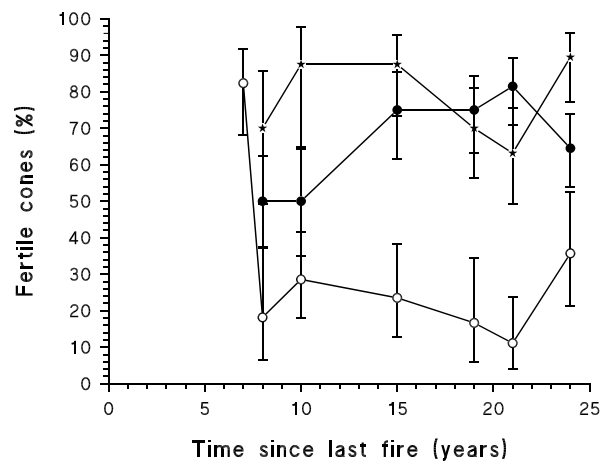

**Figure 3.** Relationship between the percentage of fertile cones (i.e. those bearing follicles) and time since the last fire (TSLF) for *B. ericifolia* (circles) and *B. marginata* (stars). Each symbol represents a population value with binomial standard errors, and the *B. ericifolia* data have been separated into the current season's cones (open symbols) and older cones (filled symbols).

**Table 2.** Average ± standard error number of fruits (follicles/bracts) per cone with respect to plant age (TSLF) and time of cone production (current/older).

| TSLF (years) | *Banksia ericifolia* | | *Banksia marginata* | | *Petrophile pulchella* | |
|---|---|---|---|---|---|---|
| | Current | Older | Current | Older | Current | Older |
| 7  | 25.9±1.9 | –        | –  | –        | 42.5±1.5  | –        |
| 8  | 22.0±1.0 | 11.7±1.6 | –  | 26.1±4.0 | 54.8±11.5 | 72.7±9.4 |
| 10 | 27.5±2.3 | 17.5±3.9 | –  | 22.7±3.4 | 51.0±8.4  | 66.5±6.0 |
| 15 | 24.5±4.9 | 28.5±2.9 | –  | 31.4±1.8 | 51.0±4.1  | 58.6±4.3 |
| 19 | 25.5±1.5 | 29.0±2.5 | –  | 38.5±5.5 | 53.3±7.7  | 55.6±4.2 |
| 21 | 39.5±4.5 | 30.5±2.8 | –  | 23.9±3.4 | 50.7±2.7  | 63.3±4.8 |
| 24 | 24.8±4.6 | 16.3±2.5 | –  | 27.9±2.0 | 64.5±9.4  | 64.8±4.7 |

**Table 3.** Results of analyses of variance testing for the effects of plant and cone ages on the average number of fruits produced per cone.

| Source of variation | *Banksia ericifolia* | | *Banksia marginata* | | *Petrophile pulchella* | |
|---|---|---|---|---|---|---|
| | F-value | $P$ | F-value | $P$ | F-value | $P$ |
| Plant age   | 2.52 | 0.025 | 2.64 | 0.031 | 0.62 | 0.713 |
| Cone age    | 2.71 | 0.102 | –    | –     | 9.25 | 0.003 |
| Interaction | 0.95 | 0.462 | –    | –     | 0.92 | 0.482 |

Older cones (i.e. those on older growth whorls) were statistically more likely to be open (i.e. have lost their seeds) for both *B. ericifolia* and *P. pulchella* but not for *B. marginata* (Table 4). However, nearly three-quarters of the oldest *P. pulchella* bracts had ruptured, while only one-quarter of the older *B. ericifolia* follicles had ruptured. *B. marginata* plants consistently had less than one-sixth of their follicles ruptured. Since the oldest cones occur on the oldest plants, this means that loss due to spontaneous fruit rupture becomes an increasingly more probable fate of the seed bank as the plants get older for *B. ericifolia* and especially *P. pulchella*.

Accumulation of total seed stored within the canopy at each site was estimated as the product of the average number of cones, percent fertile cones, average number of fruits per cone, percent of closed fruits and number of seeds per fruit. Based on the chronosequence analysis, represented by the heuristic sigmoid function, all three species accumulated a substantial seed bank over time (Fig. 4), with *B. marginata* and *P. pulchella* accumulating the largest numbers of seeds. All three species reached their maximum seed bank size 18–20 years after a fire. Ultimately, *B. marginata* accumulated a seed bank of ~3000 seeds per plant, with ~1500 seeds per plant for *P. pulchella* and ~500 for *B. ericifolia*. *B. marginata* produced a larger seed bank than the other two species by producing more cones and more seeds per fruit than they did, and by opening fewer of the fruits. These seed bank numbers do not take into account losses due to seed predation or spontaneous seed abortion for any of the species, and so they represent the potential maximum seed bank sizes.

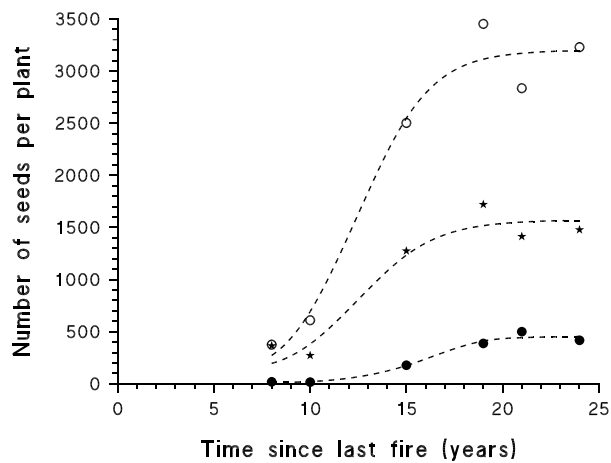

**Figure 4.** Relationship between the estimated accumulated seed bank and time since the last fire (TSLF) for *B. ericifolia* (filled circles), *B. marginata* (open circles) and *P. pulchella* (stars). Each symbol represents a population estimate, along with the predicted values (dashed lines) of the best-fit sigmoid curve (*B. ericifolia*: $r^2 = 0.977$; *B. marginata*: $r^2 = 0.976$; *P. pulchella*: $r^2 = 0.967$).

## 4. Discussion

Every organism allocates its resources to various essential activities, which can be categorised as maintenance, growth and reproduction. We examined the production of cones through time and found an exponential increase in cone number for *B. ericifolia* and *P. pulchella*. We then assessed cone production as a function of growth, with these two species showing that production of new cones was positively correlated with growth as determined by canopy volume. In terms of reduced subsequent survival, there is usually a trade-off between investment in vegetative growth and reproductive output, and this is likely to be most severe in resource poor habitats. However, the positive correlation for *B. ericifolia* and *P. pulchella* suggests that plant growth and reproduction may be allocated equal resources in these two species.

**Table 4.** Average ± standard error of the number of open and closed fruits (follicles/bracts) per cone with respect to time of cone production (measured in whorls), along with the results of the analyses of variance testing for the effect of cone ages.

| | *Banksia ericifolia* | | | *Banksia marginata* | | | *Petrophile pulchella* | |
|---|---|---|---|---|---|---|---|---|
| Whorl | Open | Closed | Whorl | Open | Closed | Whorl | Open | Closed |
| 1 | 0±0 | 26.4±1.3 | 2 | 2.1±1.3 | 29.4±3.2 | 1 | 1.3±1.3 | 52.1±3.0 |
| 2 | 0.4±0.4 | 17.9±2.2 | 3 | 0.4±0.2 | 24.9±3.1 | 2 | 2.2±1.4 | 59.2±5.9 |
| 3 | 0.7±0.5 | 20.3±3.7 | 4 | 1.6±1.0 | 29.5±5.3 | 3 | 14.9±6.1 | 53.6±5.7 |
| 4 | 2.6±1.2 | 23.4±2.8 | 5 | 2.9±1.5 | 26.1±3.0 | 4–6 | 27.7±5.6 | 33.1±4.8 |
| 5–7 | 1.8±0.7 | 21.6±2.7 | 6–8 | 2.1±0.6 | 25.9±2.2 | 7–10 | 45.6±8.0 | 16.8±6.2 |
| 8–12 | 6.9±2.1 | 19.7±3.5 | 9–14 | 5.5±5.3 | 30.5±8.0 | | | |
| F-value | 4.19 | 1.35 | F-value | 1.09 | 0.33 | F-value | 10.93 | 9.41 |
| *P* | 0.004 | 0.250 | *P* | 0.372 | 0.894 | *P* | <0.001 | <0.001 |

*B. marginata*, on the other hand, showed a weaker relationship with canopy volume, and a much more linear production of cones through time. This contrasts with the results of Zammit & Westoby (1987) and Vaughton & Ramsey (1998) who found that mean accumulation of cones was strongly related to the size of the individuals in this species, irrespective of TSLF, and may indicate that reproduction is allocated a set proportion of resources irrespective of growth. This may also be related to the fact that *B. marginata* produced twice as many cones as did either of the other two species, and did so consistently through time. Thus, either the *B. marginata* cones cost less per unit in terms of resources than do the cones of the other two species, or *B. marginata* is allocating a greater proportion of its resources to reproduction (or, perhaps, the plants are simply acquiring more resources from the same habitat).

Great variation in cone production was observed both between populations of the same TSLF age and also within populations. Cone production may thus largely be influenced by slight changes in environmental factors, as *B. ericifolia* growing on shallow soils has been reported to delay reproduction (Bradstock & O'Connell 1988). Also, natural genotypic variation among individuals will lead to greater apparent variation in older populations, as fluctuations in annual production of cones will cause a larger cumulative variation in the number of cones accumulated over time, and this will become especially evident in the older populations.

The relatively consistent position where the new cones occurred implies that each branch must reach a certain maturity before it can produce flowers. Here, *P. pulchella* primarily produced cones on whorls 2 years in age, whilst the two *Banksia* species showed production of new cones over a small range of growth whorls, with *B. marginata* producing new cones primarily on 3–4-year-old growth whorls and *B. ericifolia* producing most new cones on growth whorls 5 years in age. The positioning of cones with respect to growth whorls seemed to display a similar trend across all TSLF ages and geographical locations. However, our quantitative observations contrast with results that reported new cones on 1–3-year-old whorls in *B. ericifolia* (George 1981; Zammit & Westoby 1988). These position preferences may be related to the age required for first flowering in this species — for example, if whorls need to be > 5 years old before they can bear a cone, and one whorl is produced per year, then flowering cannot occur before this age. This position can also be used to determine a minimum age for the older cones, since the whorls are produced at the rate of one or less per year for all three species (Jenkins *et al.* 2005).

It is characteristic of Proteaceae species to produce a high percentage of barren cones (Collins & Rubelo 1987), and this has been observed in studies of several *Banksia* species (Scott 1982; Paton & Turner 1985; Whelan & Goldingay 1986; Zammit & Wood 1986). In our study 20–50 percent of all *B. ericifolia* cones produced set no follicles, whilst for *B. marginata* 10–40 percent of cones were barren, although based on observations of the current season's cones these values may be under-estimates due to loss of older barren cones. Production of follicles on only a proportion of cones suggests that resource allocation possibly out-weighs the risk of concentrating follicles, and therefore seeds, on a select number of cones that could potentially be damaged by predators such as birds and insects. In this regard it is interesting that the percentage of barren cones was much less in the *B. ericifolia* population that was producing its first major cone crop, while the average number of follicles per cone was the same as in the older populations, as this indicates a greater commitment of resources to reproduction at this time. Previous studies involving *Banksia* species suggest that the low rate of follicle production per cone could be due to a shortage of nutrients or pollinators, or to insect damage during development (Scott 1982; Paton & Turner 1985; Whelan & Goldingay 1986; Zammit & Wood 1986).

In addition, the number of follicles produced per fertile cone was observed to vary for the *Banksia* species, apparently peaking in the 15–20-year TSLF populations. This may represent a behavioural change with the age of the plants that, when combined with the increased cone production of older plants, produces a rapid increase in the potential seed bank at this time. However, this would to some extent be offset by the greater number of open follicles observed on older plants (see below). For *P. pulchella*, the 2001 season's cones had fewer bracts per cones than did the older cones, in most populations, an observation for which we offer no explanation other than potential annual variation.

In the absence of fire, seeds may be released by the spontaneous opening of old follicles or bracts, and the extent to which seeds are retained in serotinous fruits until fire or released spontaneously through time varies between and within species (Zammit & Westoby 1988). We found that spontaneous fruit rupture, and therefore seed release and loss, was related to cone age, at least for *B. ericifolia* and *P. pulchella*, with up to 75% of older fruits open in the latter and 25% in the former. This appears to be the primary cause of seed bank losses in these species.

It is usually noted that the likelihood of plant recruitment and survival to reproductive age from the seeds released in the absence of a fire is extremely small (Bradstock & O'Connell 1988; Enright *et al* 1996; Keith 1996; Whelan et al. 1998). Nevertheless, unlike the two *Banksia* species, *P. pulchella* showed some evidence of inter-fire recruitment (M. Jenkins pers. obs.), which is presumably related to the larger number of seeds released from the open bracts. Although it is suggested that recruitment is maximised due to the freeing of resources after a fire, if *P. pulchella* is one of the few species utilising the inter-fire period then some seedling survival may occur (cf. Bradstock & O'Connell 1988). We did not quantify this during our study, but further research into the proportion of individuals recruited between fires could provide further insight into the regenerative strategies, such as slow leakage of seeds as opposed to mass release of seeds after a fire event, and factors which influence seedling establishment, such as availability of light and space (Keith 1996).

Flowering is an effective indicator of reproductive maturity only if it is associated with the development of mature viable seed, and leads to the build-up of an adequate seed bank that would allow regeneration of the population in the event of a fire. Some obligate-seeder species, particularly those with woody fruits, take up to a year for their fruits to mature and are vulnerable to fire during this period — for example *B. ericifolia, Hakea teretifolia, Hakea sericea* and *P. pulchella* (Benson 1985). In our study, *B. marginata* was observed to start cone production in 7-year-old populations, whilst *B. ericifolia* and *P. pulchella* began flowering at 5 years TSLF (no 5-year-old site was sampled for *B. marginata*). However, follicle production was not observed to begin until 8 years in *B. marginata* and *B. ericifolia*, and even then extreme fluctuations in follicle production occurred.

From our results, a minimum fire-free period of greater than 10 years may be necessary for worthwhile fruit initiation in all three study species, as our estimated average time for production of a second cone on a plant, which seems to coincide with initial production of follicles/bracts, was 8.3 years for *B. ericifolia*, 6.7 years for *B. marginata* and 4.3 years for *P. pulchella*. The primary juvenile periods have been reported to be 6–8 and 5–9 years for *B. ericifolia* and *P. pulchella* respectively (Myerscough *et al.* 2000), which is consistent with our estimates. When estimating the juvenile periods of obligate-seeder species, we suggest that it is advantageous to slightly over-estimate than to under-estimate, because regeneration may not occur if erratic production of cones occurs or low fruit numbers are set in the first few years of reproduction.

Certain fire regimes are known to cause local extinction of species, as in the case when multiple fires occur during the juvenile period (Hammill *et al.* 1998). Bradstock & O'Connell (1988) found that seeds could be available at 6 years TSLF in *B. ericifolia* and 5 years in *P. pulchella*, and replacement would then be possible if seedling establishment after fire was very high. They suggested that fire-free intervals of 8–10 years would be tolerated without any population decline, but a 13-year inter-fire period would be needed for persistence if establishment of seedlings was low. Our results support these findings with respect to tolerance after 10 years TSLF, but no substantial seed production was observed before 8 years TSLF for all three species. Therefore, the minimum inter-fire period recommended in view of the present results is 8 years TSLF for all three obligate-seeder species. Even then the populations may not be able to persist if burnt at 8 years TSLF, if cone production and fruit set is low.

Furthermore, all three species would benefit from a fire-free period greater than 15 years, as reproductive potential (i.e. accumulated seed bank) seems to reach a maximum during the ages of

1318–20 years. This increasing fruit production could also be attributed to the production of extra seed as replacement for those that are damaged due to predation, or replacement for those lost/released due to spontaneous follicle or bract rupture.

Enright & Lamont (1989) reported that the quantity of seed stored in the canopy of five co-occurring Western Australian *Banksia* species varied by two orders of magnitude. Similarly, we found that the variation between the strategies and quantity of seed bank accumulated was large, as *B. marginata* was estimated to have greater than six times more accumulated seed stored than did *B. ericifolia*. This was achieved in a consistent manner across all of the aspects of the seed bank examined here, because *B. marginata* produced more cones and more seeds per fruit than the other two species did, and it opened fewer of the fruits with increasing cone age. Thus, three very different regeneration tactics were observed, highlighting the variation in reproductive strategies of co-occurring Proteaceae species.

*B. marginata* potentially accumulated a seed bank of ~3000 seeds per plant after 20 years, with ~1500 seeds per plant for *P. pulchella* and ~500 seeds per plant for *B. ericifolia*. These are clearly substantial numbers of seeds available for replacement of adults killed by a fire, but they represent maximum potential estimates, because predation and spontaneous abortion of the seeds have not been taken into account, and these are known to have important impacts on the dynamics of seed banks of Proteaceae (Scott 1982; Zammit & Wood 1986; Bradstock & O'Connell 1988; Zammit & Westoby 1988; Witkowski et al. 1991).

Ideally, the quantification of cone production and accumulation of seed bank would be done over several decades on several replicate populations; however, this takes considerable time and resources. The method of simultaneously using several populations of differing TSLF ages (i.e. a chronosequence) is a practical alternative. However, spatial differences may cause apparent temporal variation between populations, a potential problem that we sought to ameliorate by replicating TSLF ages where possible and by examining sites over a large spatial variation (~60 km). As observed in our study, this spatial variation was slight, indicating that our results are robust and are thus likely to be general for these species throughout their distribution.

Long-term canopy storage of seeds varies among species and is likely to be related to the fire regime experienced by the plant community (Whelan *et al.* 1998). We have shown that the three species commonly found together in the heath vegetation of the Sydney region have very different strategies in terms of the magnitude of serotiny and the patterns of seed accumulation with respect to TSLF. To gain an overall picture of the seed bank dynamics of obligate-seeder species, an integrated experimental study of the effects of the predation and other environmental factors on production and accumulation of seed bank and recruitment and survival after seed release will be necessary.

## Acknowledgements

Thanks to the (then) New South Wales National Parks and Wildlife Service for the use of facilities, and especially to Mark Ooi for his help. Kate Langdon and Simon Rowe helped with the fieldwork. The sampling was carried out under licences from the New South Wales National Parks and Wildlife Service and the Sydney Catchment Authority.## References

Benson, D.H. (1985) Maturation periods for fire-sensitive shrub species in Hawkesbury Sandstone vegetation. *Cunninghamia* **1**, 339–349.
Bond, W.J. (1985). Canopy-stored seed reserves (serotiny) in Cape Proteaceae. *South African Journal of Botany* **51**, 181–186.
Bradstock, R.A. & Myerscough, P.J. (1981) Fire effects on seed release and the emergence and establishment of seedlings in *Banksia ericifolia* L.f. *Australian Journal of Botany* **29**, 521–532.
Bradstock, R.A. & O'Connell, M.A. (1988) Demography of woody plants in relation to fire: *Banksia*